\documentstyle[12pt]{article}
\let\ni=\noindent

\begin{document}

\baselineskip 0.75cm
 
\pagestyle {empty}

\renewcommand{\thefootnote}{\fnsymbol{footnote}}

\newcommand{\CKM}{Cabibbo---Kobayashi---Maskawa }

\newcommand{\UK}{Super--Kamiokande }

~~~

\vspace{1.0cm}

{\large\centerline{\bf Explicit lepton texture}}

\vspace{0.8cm}

{\centerline {\sc Wojciech Kr\'{o}likowski}}

\vspace{0.6cm}

{\centerline {\it Institute of Theoretical Physics, Warsaw University}}

{\centerline {\it Ho\.{z}a 69,~~PL--00--681 Warszawa, ~Poland}}

\vspace{0.8cm}

{\centerline {\bf Abstract}}

\vspace{0.3cm}

 An explicit form of charged--lepton mass matrix, predicting $ m_\tau = 1776.80
$~MeV  from the experimental values of $ m_e $ and $ m_\mu$ (in good agreement
with the experimental figure $ m_\tau = 1777.05^{+0.29}_{-0.26}$~MeV), is 
applied to three neutrinos $\nu_e $, $\nu_\mu $, $\nu_\tau $ in order to cor\-%
relate tentatively their masses and mixing parameters. While for charged 
leptons the off--diagonal mass--matrix elements turn out to be small {\it
versus} its diagonal elements, it is suggested that for neutrinos the situation
is inverse. Under such a conjecture, the neutrino masses, lepton \CKM matrix 
and neutrino oscillation probabilities are calculated in the corresponding 
lowest (and the next to lowest) perturbative order. Then, the nearly maximal 
mixing of $\nu_\mu $ and $\nu_\tau $ is predicted in consistency with the 
observed deficit of atmospheric $\nu_\mu $'s. However, the predicted deficit of
solar $\nu_e $'s is much too small to explain the observed effect, what 
suggests the existence of (at least) one sort, $\nu_s $, of sterile neutrinos, 
whose mixing with $\nu_e$ would be responsible for the observed deficit. In the
last Section, promising perspectives for applying the same form of mass matrix 
to quarks are outlined. Two independent predictions of $|V_{ub}|/|V_{cb}| =
0.0753 \pm 0.0032 $ and unitary angle $\gamma \simeq 70^\circ $ are deduced 
from the experimental values of $|V_{us}|$ and $|V_{cb}|$ (with the use of 
quark masses $ m_s $, $ m_c $ and $ m_b $).

\vspace{0.3cm} 

\ni PACS numbers: 12.15.Ff , 14.60.Pq , 12.15.Hh 
 
\vspace{1.5cm} 

\ni November 1998

\vfill\eject

\pagestyle {plain}

\setcounter{page}{1}

~~~

\vspace{0.2cm}

\ni {\bf 1. Introduction}

\vspace{0.3cm}

 In this paper, the explicit form of mass matrix invented for three generations
of charged leptons $ e^-\,,\,\mu^-\,$, $\tau^- $, and being surprisingly good
for their masses [1], is applied to three generations of neutrinos $\nu_e $, $
\nu_\mu $, $\nu_\tau $, in order to correlate tentatively their masses and
mixing parameters. This form reads

\vspace{-0.3cm}

\begin{equation}
\left({M}^{(f)}_{ij}\right) = \frac{1}{29} \left(\begin{array}{ccc} 
\mu^{(f)}\varepsilon^{(f)} & 2\alpha^{(f)} e^{i\varphi^{(f)}} & 0 \\ & & 
\\ 2\alpha^{(f)} e^{-i\varphi^{(f)}} & 4\mu^{(f)}(80 + \varepsilon^{(f)})/9 
& 8\sqrt{3}\,\alpha^{(f)} e^{i\varphi^{(f)}} \\ & & 
\\ 0 & 8\sqrt{3}\,\alpha^{(f)} e^{-i\varphi^{(f)}} & 24\mu^{(f)}
(624 + \varepsilon^{(f)})/25 \end{array}\right) \;,
\end{equation}

\vspace{-0.1cm}

\ni where the label $f = \nu\,,\,e$ denotes neutrinos and charged leptons, 
respecively, while $\mu^{(f)}$, $\varepsilon^{(f)}$, $\alpha^{(f)}$ and 
$\varphi^{(f)}$ are real constants to be determined from the present and 
future experimental data for lepton masses and mixing parameters ($\mu^{(f)}$
and $\alpha^{(f)}$ are mass--dimensional). In our approach, neutrinos are 
assumed to carry pure Dirac masses.

 Here, the form (1) of mass matrices $\left({M}^{(\nu)}_{ij}\right)$ and $
\left({M}^{(e)}_{ij}\right)$ may be considered as a detailed ansatz to be 
compared with the lepton data. However, in the past, we have presented an 
argument [2,1] in favour of the form (1), based on: ({\it i}) K\"{a}hler--like
generalized Dirac equations (interacting with the Standard--Model gauge bosons)
whose {\it a priori} infinite series is necessarily reduced (in the case of 
fermions) to three Dirac equations, due to an intrinsic Pauli principle, and
({\it ii}) an ansatz for the fermion mass matrix, suggested by the above three%
--generation characteristics ({\it i}).

 In the case of charged leptons, assuming that the off--diagonal elements of 
the mass matrix $\left({M}^{(e)}_{ij}\right)$ can be treated as a small 
perturbation of its diagonal terms ({\it i.e.}, that $\alpha^{(e)}/\mu^{(e)}$ 
is small enough), we calculate in the lowest perturbative order [1]

\vspace{-0.3cm}

\begin{eqnarray}
m_\tau & = & \left[ 1776.80 + 10.2112 \left(\frac{\alpha^{(e)}}{\mu^{(e)}}
\right)^2\,\right]\;{\rm MeV} \;, \nonumber \\
\mu^{(e)} & = & 85.9924\;{\rm MeV} + O\left[\left(\frac{\alpha^{(e)}}{\mu^{(e)}
}\right)^2\,\right]\,\mu^{(e)} \;,\nonumber \\
\varepsilon^{(e)} & = & 0.172329 + O\left[\left(\frac{\alpha^{(e)}}{\mu^{(e)
}}\right)^2 \right]\;,
\end{eqnarray}

\vspace{-0.1cm}

\ni when the experimental values of $ m_e $ and $ m_\mu $ [3] are used as 
inputs. In Eqs. (2), the first terms are given as $\stackrel{\circ}{m}_\tau =
6(351m_\mu - 136 m_e)/125 $, $\stackrel{\circ}{\varepsilon}^{(e)} = 320
m_e/(9m_\mu - 4 m_e)$ and $\stackrel{\circ}{\mu}^{(e)} = 29(9m_\mu - 4 m_e)
/320 $, respectively. We can see that the predicted value of $ m_\tau $ agrees 
very well with its experimental figure $ m_\tau^{\rm exp} = 1777.05^{+0.29}_{
-0.26}$~MeV [3], even in the zero perturbative order. To estimate 
$\left(\alpha^{(e)}/\mu^{(e)}\right)^2 $, we can take this experimental figure 
as another input, obtaining

\vspace{-0.2cm}

\begin{equation}
\left(\frac{\alpha^{(e)}}{\mu^{(e)}}\right)^2 = 0.024^{+0.028}_{-0.025} \;,
\end{equation}

\vspace{-0.1cm}

\ni  which value is not inconsistent with zero. Hence, $\alpha^{(e)\,2} = 180
^{+210}_{-190}\,{\rm MeV}^2 $ due to Eq. (2).

 For the unitary matrix $\left({U}^{(e)}_{ij}\right)$, diagonalizing the 
charged--lepton mass matrix $\left({M}^{(e)}_{ij}\right)$ according to the 
relation $ U^{(e)\,\dagger}\,M^{(e)}\,U^{(e)} = {\rm diag}(m_e\,,\,m_\mu\,,\,
m_\tau)$, we get in the lowest perturbative order 

\vspace{-0.25cm}

\begin{equation}
\left( U^{(e)}_{ij}\right) = \left(\begin{array}{ccc} 1 - \frac{2}{29^2}\left(
\frac{\alpha^{(e)}}{m_\mu}\right)^2 & \frac{2}{29} \frac{\alpha^{(e)}}{m_\mu}
e^{i\varphi^{(e)}} & \frac{16\sqrt{3}}{29^2} \left(\frac{\alpha^{(e)}}{m_\tau} 
\right)^2 e^{2i \varphi^{(e)}} \\ & & \\ -\frac{2}{29}\frac{\alpha^{(e)}}
{m_\mu} e^{-i\varphi^{(e)}} & 1 - \frac{2}{29^2}\left(\frac{\alpha^{(e)}}{
m_\mu}\right)^2 - \frac{96}{29^2}\left(\frac{\alpha^{(e)}}{m_\tau}\right)^2 & 
\frac{8\sqrt{3}}{29}\frac{\alpha^{(e)}}{m_\tau}e^{i\varphi^{(e)}} \\  & & \\
\frac{16\sqrt{3}}{29^2}\frac{\alpha^{(e)\,2}}{m_\mu\,m_\tau}\,e^{-
2i \varphi^{(e)}} & -\frac{8\sqrt{3}}{29}\frac{\alpha^{(e)}}{m_\tau} e^{-i
\varphi^{(e)}} & 1 - \frac{96}{29^2}\left(\frac{\alpha^{(e)}}{m_\tau}\right)^2 
\end{array} \right) \;.
\end{equation}

\vspace{0.2cm}

\ni {\bf 2. Neutrino masses and mixing parameters}

\vspace{0.3cm}

 In the case of neutrinos, because of their expected tiny mass scale $\mu^{(\nu
)}$, we will tentatively conjecture that the diagonal elements of the mass 
matrix $\left( M^{(\nu)}_{ij}\right)$ can be treated as a small perturbation of
its off--diagonal terms ({\it i.e.}, that $\mu^{(\nu)}/\alpha^{(\nu)}$ is small
enough). In addition, we put $\varepsilon^{(\nu)} = 0 $ {\it i.e.}, $ M^{(
\nu)}_{11} = 0 $. Then, we calculate in the lowest perturbative order the 
following neutrino masses:

\vspace{-0.3cm}

\begin{eqnarray}
m_{\nu_1} & = & \frac{|M^{(\nu)}_{12}|^2\,M^{(\nu)}_{33}}{|M^{(\nu)}_{12}|^2
+ |M^{(\nu)}_{23}|^2} = \frac{1}{49}M^{(\nu)}_{33}  = \frac{1}{49} \xi |M^{(\nu
)}_{12}|\;, \nonumber \\ 
m_{\nu_2,\,\nu_3} & = &  \mp \sqrt{|M^{(\nu)}_{12}|^2 + |M^{(\nu)}_{23}|^2}
+ \frac{1}{2}\left( \frac{48}{49}M^{(\nu)}_{33} + M^{(\nu)}_{22}\right)
\nonumber \\ & = & \left[ \mp 7 + \frac{1}{2}\left( \frac{48}{49} \xi + \chi
\right)\right] |M^{(\nu)}_{12}| \;,
\end{eqnarray}

\vspace{-0.1cm}

\ni where

\vspace{-0.2cm}

\begin{eqnarray}
\xi & \equiv & \frac{M^{(\nu)}_{33}}{|M^{(\nu)}_{12}|}= \frac{7488}{25}\frac{
\mu^{(\nu)}}{\alpha^{(\nu)}} = 299.52\frac{\mu^{(\nu)}}{\alpha^{(\nu)}}
\;, \nonumber \\ 
\chi & \equiv & \frac{M^{(\nu)}_{22}}{|M^{(\nu)}_{12}|}= \frac{160}{9}\frac{
\mu^{(\nu)}}{\alpha^{(\nu)}} = \frac{125}{2106} \xi = \frac{1}{16.848} \xi \;,
\end{eqnarray}

\vspace{-0.1cm}

\ni are relatively small by our perturbative conjecture, while

\vspace{-0.1cm}

\begin{equation}
|M^{(\nu)}_{12}| = \frac{2}{29}\alpha^{(\nu)}\;,\; |M^{(\nu)}_{23}| = 
\frac{8\sqrt{3}}{29}\alpha^{(\nu)} =\sqrt{48} |M^{(\nu)}_{12}| \;.
\end{equation}

\vspace{-0.1cm}

\ni As seen from Eqs. (5), the actual perturbative parameters are not $\xi $ 
and $\chi $, but rather $\xi/7 $ and $\chi/7 $, what is confirmed later in Eqs.
(9). Note that $ m_{\nu_2} < 0 $, the minus sign being irrelevant in the 
relativistic case, where only $ m_{\nu_2}^2 $ is measured ({\it cf.} Dirac 
equation): $ |m_{\nu_2}| $ may be considered as a phenomenological mass of $
\nu_2 $.

 Using Eqs. (5), we can write the formula

\vspace{-0.1cm}

\begin{equation}
m_{\nu_3}^2 - m_{\nu_2}^2 = 14\left(\frac{48}{49}\xi + \chi\right)|M^{(\nu)}_{
12}|^2 = 20.721\, \alpha^{(\nu)} \mu^{(\nu)} \;,
\end{equation}

\vspace{-0.1cm}

\ni which will enable us to determine the product  $\alpha^{(\nu)}\mu^{(\nu)}$ 
from the observed deficit of atmospheric neutrinos $\nu_\mu $, if $\nu_\mu 
\rightarrow \nu_\tau $ oscillations are really responsible for this effect.

 We calculate also the unitary matrix $\left(U_{ij}^{(\nu)}\right)$ diagonal\-%
izing  the neutrino mass matrix $\left(M_{ij}^{(\nu)}\right)$ according to
the relation $ U^{(\nu)\,\dagger} M^{(\nu)} U^{(\nu)} = {\rm diag}(m_{\nu_1}\,
,\,m_{\nu_2}\,,\,m_{\nu_3})$. In the lowest perturbative order we obtain

\vspace{-0.2cm}

\begin{eqnarray}
U^{(\nu)}_{11} & = & \sqrt{ \frac{48}{49} }\left[1 - \left(\frac{24}{49^3} -
\frac{1}{49^4}\right)\,\xi^2\right] \;, \nonumber \\
U^{(\nu)}_{21} & = & \frac{1}{49}\sqrt{\frac{48}{49}} \xi e^{-i\varphi^{(\nu)}
} \;, \nonumber \\
U^{(\nu)}_{31} & = & -\frac{1}{7}\left[1 - \left(\frac{73}{49^3} - 
\frac{1}{49^4}\right) \xi^2 + \frac{1}{49}\xi \chi \right] e^{-2i 
\varphi^{(\nu)}} \;, \nonumber \\
U^{(\nu)}_{12} & = & -\frac{1}{\sqrt{2}} \frac{1}{7} \left(1 +
\frac{36}{7\cdot 49} \xi + \frac{1}{28}\chi \right)\,e^{i\varphi^{(\nu)}}
\;, \nonumber \\
U^{(\nu)}_{22} & = & \frac{1}{\sqrt{2}} \left(1 + \frac{12}{7\cdot 49} \xi 
- \frac{1}{28}\chi \right) \;, \nonumber \\
U^{(\nu)}_{32} & = & -\frac{1}{\sqrt{2}} \sqrt{\frac{48}{49}} \left(1 -
\frac{13}{7\cdot 49} \xi + \frac{1}{28} \chi \right)\,e^{-i \varphi^{(\nu)}}
\;, \nonumber \\
U^{(\nu)}_{13} & = & \frac{1}{\sqrt{2}} \frac{1}{7} \left(1 -
\frac{36}{7\cdot 49} \xi - \frac{1}{28}\chi \right)\,e^{2i \varphi^{(\nu)}}
\;, \nonumber \\
U^{(\nu)}_{23} & = & \frac{1}{\sqrt{2}} \left(1 - \frac{12}{7\cdot 49} \xi + 
\frac{1}{28}\chi \right)\,e^{i\varphi^{(\nu)}} \;, \nonumber \\
U^{(\nu)}_{33} & = & \frac{1}{\sqrt{2}} \sqrt{\frac{48}{49}} \left(1 +
\frac{13}{7\cdot 49} \xi - \frac{1}{28} \chi \right) 
\end{eqnarray}

\vspace{-0.1cm}

\ni with $\chi = (125/2106)\xi = \xi/16.848 $.

 Denoting by $\nu_\alpha = \nu_e\,,\,\nu_\mu \,,\,\nu_\tau $ and $\nu_i = 
\nu_1\,,\,\nu_2\,,\,\nu_3 $ the weak--interaction and mass neutrino fields, 
respectively, we have the unitary transformation

\vspace{-0.2cm}

\begin{equation}
\nu_\alpha = \sum_i \left( V^\dagger \right)_{\alpha\,i}\nu_i = \sum_i V^*_{i\,
\alpha} \nu_i \;,
\end{equation}

\vspace{-0.1cm}

\ni where the lepton counterpart $\left( V_{\alpha\,i}\right) $ of the \CKM 
matrix is given as $ V = U^{(\nu)\,\dagger} U^{(e)} \simeq U^{(\nu)\,\dagger}$ 
or 

\vspace{-0.2cm}

\begin{equation}
V_{i\,\alpha} = \sum_\beta \left( U^{(\nu)\,\dagger} \right)_{i\,\beta}
U^{(e)}_{\beta\,\alpha} \simeq U^{(\nu)\,*}_{\alpha\,i} \;,
\end{equation}

\vspace{-0.1cm}

\ni the approximate equality being valid for negligible $\alpha^{(e)}/\mu^{(e)}
$ when $ U_{\beta\,\alpha}^{(e)} \simeq \delta_{\beta\,\alpha}$ due to Eq. (4).
Of course, in Eqs. (9) we wrote $\alpha = 1,2,3 $ for simplicity. From Eq. 
(10), we get the unitary transformation $ |\nu_\alpha\rangle = \sum_i |\nu_i
\rangle V_{i\,\alpha}$, where $ |\nu_\alpha\rangle = \nu^\dagger_\alpha |0 
\rangle $ and $ |\nu_i\rangle = \nu^\dagger_i |0 \rangle $ are weak--inter\-%
action and mass neutrino states.

 In the limit of $\mu^{(\nu)} \rightarrow 0 $ (implying $\xi \rightarrow 0 $ 
and $\chi \rightarrow 0 $), we obtain from Eqs. (10), (11) and (9) the follow%
ing unperturbed mixing formulae for $\nu_1\,,\,\nu_2\,,\,\nu_3 $ :

\vspace{-0.3cm}

\begin{eqnarray}
\nu_e  & \rightarrow & \frac{1}{7} \left[\sqrt{48} \nu_1 e^{-i \varphi^{(\nu)
}} - \frac{1}{\sqrt{2}}\left(\nu_2 - \nu_3 e^{i \varphi^{(\nu)}}\right)\right]
e^{i \varphi^{(\nu)}} \;, \nonumber \\ 
\nu_\mu & \rightarrow & \frac{1}{\sqrt{2}}\left(\nu_2 + \nu_3 e^{i \varphi^{(
\nu)}} \right) \;, \nonumber \\ 
\nu_\tau & \rightarrow &  -\frac{1}{7} \left[\nu_1 e^{-i \varphi^{(\nu)}} +
\sqrt{\frac{48}{2}}\left(\nu_2 - \nu_3 e^{i \varphi^{(\nu)}}\right)\right]
e^{-i \varphi^{(\nu)}} \;.
\end{eqnarray}

\vspace{-0.1cm}

\ni These display the maximal mixing between $\nu_2 $ and $\nu_3 $ in all three
cases and a smaller mixing of $\left[\nu_2 - \nu_3 \exp \left(i\varphi^{(\nu)}
\right)\right]/\sqrt{2}$ with $\nu_1 $ in the cases of $\nu_e$ and $\nu_\tau $, 
giving a minor admixture to $\nu_e $ and a dominating admixture to $\nu_\tau $ 
(in $\nu_\mu $ there is no admixture of $\nu_1 $).

\vspace{0.3cm}

\ni {\bf 3. Neutrino oscillations}

\vspace{0.3cm}

 Once knowing the elements $ V_{i\,\alpha}$ of the lepton \CKM matrix, we can 
calculate the probabilities of neutrino oscillations $\nu_\alpha \rightarrow 
\nu_\beta$ (in the vacuum) making use of the general formula

\vspace{0.1cm}

\begin{equation}
P(\nu_\alpha \rightarrow \nu_\beta) = |\langle\nu_\beta |\nu_\alpha(t)\rangle
|^2 = \sum_{i\,j}V_{j\,\beta}V^*_{j\,\alpha}V^*_{i\,\beta}V_{i\,\alpha} 
e^{2i x_{j\,i}} \;,
\end{equation}

\vspace{0.1cm}

\ni where $|\nu_\alpha(t)\rangle = \exp(-i H t) |\nu_\alpha \rangle $ and 


\begin{equation}
x_{j\,i} = 1.26693\, \Delta m^2_{j\,i}\, L/E \;\;,\;\;\Delta m^2_{j\,i} = 
m^2_{\nu_j} - m^2_{\nu_i}\;\;,
\end{equation}


\ni if $\Delta m^2_{j\,i}$, $ L $ and $ E $ are measured in eV$ ^2 $, km and 
GeV, respectively, with $ L = t $ and $ E = |\vec{p}|\;\;(c = 1 = \hbar) $ 
denoting the experimental baseline and neutrino energy.

 It is not difficult to show that for the mass matrix $\left( M^{(\nu)}_{ij}
\right)$, as it is given in Eq. (1), the quartic products of $V_{i\,\alpha}$'s 
in Eq. (13) are always real (for any phase $\varphi^{(\nu)}$), if only $V_{i\,
\alpha} = U^{(\nu)\,*}_{\alpha\,i}$ ({\it i.e.}, $ U^{(e)}_{\beta\,\alpha} = 
\delta_{\beta\,\alpha}$). This implies that $P(\nu_\alpha \rightarrow \nu_\beta
) = P(\nu_\beta \rightarrow \nu_\alpha)$. In general, the last relation is valid
in the case of CP invariance which, under the CPT theorem, provides the time--%
reversal invariance. Because of the real values of quartic products of $V_{i\,
\alpha}$'s, the formula (13) can be rewritten as


\begin{equation}
P(\nu_\alpha \rightarrow \nu_\beta) = \delta_{\beta\,\alpha} - 4 \sum_{i<j}
V_{j\,\beta}V^*_{j\,\alpha}V^*_{i\,\beta}V_{i\,\alpha} \sin^2 x_{j\,i} 
\end{equation}


\ni without the necessity of introducing phases of these products.

 With the lowest--order perturbative expressions (9) for $V_{i\,\alpha} = U^{(
\nu)\,*}_{\alpha\,i}$, the formula (15) leads to the following forms of 
appearance oscillation probabilities:


\begin{eqnarray}
\lefteqn{P\left(\nu_\mu \rightarrow \nu_e  \right) = \frac{1}{49}\sin^2 x_{32}}
\nonumber \\ &  & + \frac{96}{7\cdot 49^2} \xi \left[\left(1 + \frac{48}{7
\cdot 49} \xi \right) \sin^2 x_{21} - \left(1 - \frac{48}{7\cdot 49} \xi\right)
\sin^2 x_{31}\right] \;,\\
\lefteqn{P\left(\nu_\mu \rightarrow \nu_\tau\right) = \frac{48}{49}\sin^2 x_{32
}} \nonumber \\ &  & + \frac{96}{7\cdot 49^2} \xi \left[ -\left(1 - \frac{1}{
7\cdot 49} \xi \right) \sin^2 x_{21} + \left(1 + \frac{1}{7\cdot 49} \xi\right)
\sin^2 x_{31}\right] \;,\\
\lefteqn{P\left(\nu_e \rightarrow \nu_\tau  \right) = - \frac{48}{49^2}
\sin^2 x_{32}} \nonumber \\ &  & + \frac{96}{49^2}\left[\left(1 + \frac{23}{7
\cdot 49} \xi + \frac{1}{14} \chi \right) \sin^2 x_{21} + \left(1 - \frac{23
}{7\cdot 49} \xi - \frac{1}{14} \chi \right) \sin^2 x_{31}\right] 
\end{eqnarray}


\ni as well as of survival oscillation probabilities :


\begin{eqnarray}
\lefteqn{P\left(\nu_e \rightarrow \nu_e  \right) = 1 - \frac{1}{49^2}
\sin^2 x_{32}} \nonumber \\ &  & - \frac{96}{49^2}\left[\left(1 + \frac{72}{7
\cdot 49} \xi + \frac{1}{14} \chi\right) \sin^2 x_{21} + \left(1 - \frac{72}{7
\cdot 49} \xi - \frac{1}{14} \chi\right)\sin^2 x_{31}\right] \;,\\
\lefteqn{P\left(\nu_\mu \rightarrow \nu_\mu\right) = 1 - \sin^2 x_{32}
- \frac{96}{49^3} \xi^2 \left(\sin^2 x_{21} + \sin^2 x_{31}\right) \;,} \\
\lefteqn{P\left(\nu_\tau \rightarrow \nu_\tau\right) = 1 - \left(\frac{48}{49}
\right)^2\sin^2 x_{32}} \nonumber \\ &  & - \frac{96}{49^2}\left[\left(1 -
\frac{26}{7\cdot 49} \xi + \frac{1}{14} \chi \right) \sin^2 x_{21} + \left(
1 + \frac{26}{7\cdot 49} \xi - \frac{1}{14} \chi \right) \sin^2 x_{31}\right] 
\;.
\end{eqnarray}

\ni Thus, we get $ P\left(\nu_e \rightarrow \nu_e \right) + P\left(\nu_e 
\rightarrow \nu_\mu\right) + P\left(\nu_e \rightarrow \nu_\tau\right) = 1 $ and
two other obvious summation rules for probabilities. Among these probabilities,
$P\left(\nu_\mu \rightarrow \nu_\mu\right)$ displays (in the lowest perturbat%
ive order) maximal mixing between $\nu_2 $ and $\nu_3 $.

 In the lowest perturbative order,


\begin{equation}
x_{31} - x_{21} = x_{32} = 14\left(\frac{48}{49} \xi + \chi \right)
\left(1.26693 |M^{(\nu)}_{12}|\, L/E \right)
\end{equation}

\ni due to Eqs. (8) and (14). Hence,


\begin{equation}
\sin^2 x_{31} = \sin^2 x_{21} + x_{32}\sin 2x_{21} + x^2_{32}\sin 2x_{21}
\end{equation}

\ni in experiments where $ x_{32} \ll \pi/2 $. When in such cases the relation
(23) is inserted into the formulae (16), (17) and (20), its $ x_{32}$ and 
$ x^2_{32}$ terms can be neglected in the lowest perturbative order.

 Note that the mass formulae (5) imply $m^2_{\nu_1} \ll m^2_{\nu_2} \stackrel{
_<}{_\sim} m^2_{\nu_3} $, where $m^2_{\nu_1}/m^2_{\nu_2,\nu_3} = \xi^2/49^3 
+ O(\xi^3)$ and $m^2_{\nu_2}/m^2_{\nu_3} = 1 -(2/7)(48 \xi/49 + \chi) + O(
\xi^3)$. Thus, the inequality $x_{31} \stackrel{_>}{_\sim} x_{21} \gg x_{32}
$ holds in all neutrino oscillation experiments (with some given $ L $ and 
$ E $).

 We have calculated the neutrino masses, lepton \CKM matrix and neutrino oscil%
lation probabilities also in the next to lowest perturbative order. Then, in 
Eqs. (5) the mass $m_{\nu_1}$ gets no quadratic correction, while $m_{\nu_2}$ 
and $m_{\nu_3}$ are corrected by the terms

\begin{equation}
\mp \frac{1}{14}\left(\frac{13\cdot 48}{49^2} \xi^2 - \frac{24}{49} \xi \chi
+ \frac{1}{4} \chi^2\right)|M^{(\nu)}_{12}| \;,
\end{equation}

\ni respectively. Among the derived oscillation formulae, Eq. (20), for 
instance, is extended to the form

\vspace{-0.2cm}

\begin{eqnarray}
\lefteqn{P\left(\nu_\mu \rightarrow \nu_\mu\right) = 1 - \left(1 - 
\frac{672}{49^3} \xi^2 + \frac{24}{49^2} \xi \chi - \frac{1}{4\cdot 49} 
\chi^2 \right) \sin^2 x_{32} } \nonumber \\ & & \;\;\;\;\;\;\;\;\;\;\;
\;\;\;\;\;\;\;\;\;\; - \frac{96}{49^3} \xi^2 \left( \sin^2 x_{21} + 
\sin^2 x_{31}\right) \nonumber \\ 
& &\;\;\;\;\;\;\;\;\;\;\;\;\;\; = 1 - \left(1 - 0.00514\,\xi^2 \right)\sin^2 
x_{32} - 0.000816\,\xi^2 \left(\sin^2 x_{21} + \sin^2 x_{31}\right)\;\;\;\;\;
\end{eqnarray}

\vspace{-0.1cm}

\ni displaying nearly maximal mixing between $\nu_2 $ and $\nu_3 $. 

In the case of \UK atmospheric neutrino experiment [4], if $\nu_\mu \rightarrow
\nu_\tau $ oscillations are responsible for the observed deficit of atmospheric
$\nu_\mu $'s, we have $ x_{\rm atm} = x_{32} \ll x_{21} \stackrel{_<}{_\sim} 
x_{31}$, what implies that $\sin^2 x_{21} = \sin^2 x_{31} = 1/2 $ due to 
averaging over many oscillation lengths. Then, Eq. (25) leads to the following 
effective two--flavor oscillation formula:

\vspace{-0.2cm}

\begin{equation}
P\left(\nu_\mu \rightarrow \nu_\mu \right) = 1 - \left(1 - 0.00350\,\xi^2
\right) \sin^2 x_{32}\;,
\end{equation}

\ni if we assume in Eq. (25) that $0.000816 \xi^2 = 0.000816 \xi^2 (2\sin^2 
x_{32})$ effectively. Identifying the estimation (26) with the two--flavor formula
fitted in the \UK experiment, we obtain the limits

\vspace{-0.2cm}

\begin{eqnarray}
1 - 0.00350\,\xi^2 & \equiv & \sin^2 2\theta_{\rm atm} \sim 0.82\;\;{\rm to}
\;\; 1 \;, \nonumber \\ \Delta m_{32}^2 & \equiv & \Delta m_{\rm atm}^2 \sim 
(0.5\;\;{\rm to}\;\;6)\times 10^{-3}\,{\rm eV}^2 \;.
\end{eqnarray}

\ni Hence, $\xi \sim 7.17$ to 0 and

\vspace{-0.2cm}

\begin{eqnarray}
\frac{\mu^{(\nu)}}{\alpha^{(\nu)}} & \equiv & 0.00334 \xi \sim 0.0239\;\;{\rm 
to}\;\;0 \;, \nonumber \\ \alpha^{(\nu)}\mu^{(\nu)} & \equiv & 0.483\Delta m_{32
}^2 \sim (0.241\;\;{\rm to}\;\;2.90)\times 10^{-4}\,{\rm eV}^2 \;,
\end{eqnarray}

\ni where Eqs. (6) and (8) are used. For instance, with $\sin^2 2\theta_{\rm 
atm} \sim 0.999 $ and $\Delta m^2_{\rm atm} \sim 5\times 10^{-3}\,{\rm eV}^2 $,
we get $\xi \sim 0.535 $ and

\begin{equation}
\frac{\mu^{(\nu)}}{\alpha^{(\nu)}} \sim 0.00178 \;\;,\;\; \alpha^{(\nu)}
\mu^{(\nu)} \sim 2.41\times 10^{-4}\,{\rm eV}^2 \;,
\end{equation}

\ni what gives the estimation

\vspace{-0.2cm}

\begin{equation}
\alpha^{(\nu)} \sim 0.368\,{\rm eV}\;\;,\;\;\mu^{(\nu)} \sim 6.55\times 10^{-4}
\,{\rm eV} \;.
\end{equation}

\ni Note that $ \xi < 1 $ for $\sin^2 2\theta_{\rm atm} > 0.9965 $. As was 
already mentioned, our actual perturbative parameters are not $\xi $ and $\chi 
$, but rather $\xi/7 $ and $\chi/7 = 0.0594 \xi/7 $.

 Having estimated $\alpha^{(\nu)}$ and $\mu^{(\nu)}$, we can calculate neutrino
masses from Eqs. (5) with (6) and (7). Making use of the values (30) (valid for
$\sin^2 2\theta_{\rm atm} \sim 0.999 $ and $\Delta m^2_{\rm atm} \sim 5\times 
10^{-3}\,{\rm eV}^2 $), we obtain

\vspace{-0.2cm}

\begin{equation}
m_{\nu_1} \sim 2.76\times 10^{-4}\,{\rm eV} \;\;,\;\;m_{\nu_2} \sim -1.71
\times 10^{-1}\,{\rm eV} \;\;,\;\;m_{\nu_3} \sim 1.85\times 10^{-1}\,{\rm eV}.
\end{equation}

\ni Because of the smallness of these masses, the neutrinos $\nu_1 $, $\nu_2 $,
$\nu_3 $ are not likely to be responsible for the entire hot dark matter.

 In the case of solar neutrino experiments, all three popular fits [5] of the 
observed deficit of solar $\nu_e $'s to an effective two--flavor oscillation 
formula require $\Delta m^2_{\rm sol} \ll \Delta m^2_{\rm atm}$ what implies $
\Delta m^2_{\rm sol}\ll \Delta m^2_{32} \ll \Delta m^2_{21} \stackrel{_<}{_\sim}
\Delta m^2_{31}$, if $\nu_\mu \rightarrow \nu_\tau $ oscillations are res\-%
ponsible for the deficit of atmospheric $\nu_\mu $'s. Then, $ x_{\rm sol} \ll 
x_{32} \ll x_{21} \stackrel{_<}{_\sim} x_{31}$, giving $\sin^2 x_{32} = \sin^2 
x_{21} = \sin^2 x_{31} = 1/2 $ due to averaging over many oscillation lengths.
In such a case, Eq. (19) leads to

\vspace{-0.2cm}

\begin{equation}
P\left(\nu_e \rightarrow \nu_e \right) = 1 - \frac{193}{2\cdot 49^2} = 1 -
0.0402 = 0.960 \;,
\end{equation}

\ni predicting only a 4\% deficit of solar $\nu_e$'s, much too small to explain
solar neutrino observations.

 An intriguing situation arises in the case of formula (16) for $ P\left(
\nu_\mu \rightarrow \nu_e \right)$, if $\nu_\mu \rightarrow \nu_\tau $ oscil%
lations really cause the bulk of deficit of atmospheric $\nu_\mu $'s. Then, 
for a new $ x_{\rm new} = x_{32} \ll x_{21} \stackrel{_<}{_\sim} x_{31}$ (with 
some new $ L $ and $ E $) we may have $\sin^2 x_{21} = \sin^2 x_{31} = 1/2$ due
to averaging over many oscillation lengths and so, infer from Eq. (16) that

\vspace{-0.2cm}

\begin{equation}
P\left(\nu_\mu \rightarrow \nu_e \right) = \frac{1}{49} \sin^2 x_{32} + \frac{2
\cdot 48^2}{49^4} \xi^2 \sim 0.0204 \sin^2 x_{32} + 2.29\times 10^{-4}\;,
\end{equation}

\ni where $\xi^2 \sim 0.286 $ (what is valid for $\sin^2 2\theta_{\rm atm} \sim
0.999 $ and $\Delta m^2_{\rm atm} \sim 5\times 10^{-3}\,{\rm eV}^2 $). Such a 
predicted oscillation amplitude $\sin^2 2\theta_{\rm new} \sim 0.02 $ would lie
in the range of $\sin^2 2\theta_{\rm LSND}$ estimated in the positive (though 
still requiring confirmation) LSND accelerator experiment on $\nu_\mu 
\rightarrow \nu_e $ oscillations [6]. However, the lower limit $\Delta m^2_{
\rm LSND} \stackrel{_>}{_\sim} 0.1\,{\rm eV}^2 $ reported by this experiment 
is by one order of magnitude larger than the \UK upper limit $\Delta
m^2_{32} \stackrel{_<}{_\sim} 0.01\,{\rm eV}^2 $. On the other hand, the small 
predicted oscillation amplitude $\sin^2 2\theta_{\rm new} \sim 0.02 $ would not
be in conflict with the negative result of the CHOOZ long--baseline reactor 
experiment on $\bar{\nu}_e \rightarrow \bar{\nu}_\mu $ oscillations [7].

 In conclusion, our explicit model of lepton texture displays a number of 
important features. {\it (i)} It correlates correctly (with high precision) the
tauon mass with electron and muon masses. {\it (ii)} It predicts (without para%
meters) the maximal mixing between muon and tauon neutrinos in the limit 
$\mu^{(\nu)} \rightarrow 0 $, consistent with the observed deficit of atmosphe%
ric $\nu_\mu $'s. {\it (iii)} It fails to explain the observed deficit of solar
$\nu_e $'s. {\it (iv)} It predicts new $\nu_\mu \rightarrow \nu_e$ oscillations
with the amplitude consistent with LSND experiment, but with a phase corres%
ponding to the mass squared difference at least one order of magnitude smaller.

 In the framework of our model, the point {\it (iii)} may suggest that in 
Nature there exists (at least) one sort, $\nu_s $, of sterile neutrinos (blind 
to the Standard--Model interactions), responsible for the observed deficit of 
solar $\nu_e $'s through $\nu_e \rightarrow \nu_s $ oscillations dominating the
survival probability $P(\nu_e \rightarrow \nu_e) \simeq 1 - P(\nu_e \rightarrow
\nu_s)$. In an extreme version of this pict\-ure, it might even happen that in 
Nature there would be two sorts, $\nu_s $ and $\nu'_s $, of sterile neutrinos, 
where $\nu'_s $ would replace $\nu_\tau $ in explaining the observd deficit of 
atmospheric $\nu_\mu $'s by means of $\nu_\mu \rightarrow \nu'_s $ oscillations
that should dominate the survival probability $P(\nu_\mu \rightarrow \nu_\mu) 
\simeq 1 - P(\nu_\mu \rightarrow \nu'_s)$. In this case, the masses of active 
neutrinos might be even zero, what would correspond to $\alpha^{(\nu)} = 0 $ 
and $\mu^{(\nu)} = 0 $ giving then $\nu_e = \nu_1\;,\;\nu_\mu = \nu_2\;,\;
\nu_\tau = \nu_3 $ (however, very small $\alpha^{(\nu)}$ and $\mu^{(\nu)}$
leading to tiny masses and mixings of $\nu_1\;,\;\nu_2\;,\;\nu_3 $ would be 
still allowed).

 For the author of the present paper the idea of existence of two sorts of 
sterile neutrinos is fairly appealing, since two such spin--1/2 fermions, blind 
to all Standard--Model interactions, do follow (besides three standard families
of active leptons and quarks) [8] from the argument {\it (i)} mentioned in 
Introduction, based on the K\"{a}hler--like generalized Dirac equations. Note
in addition that the $\nu_e \rightarrow \nu_s $ and $\nu_\mu \rightarrow \nu'_s
$ oscillations caused by appropriate mixings should be a natural consequence 
of the spontaneous breaking of electroweak $ SU(2)\times U(1) $ symmetry.

\vspace{0.2cm}

\ni {\bf 4. Perspectives for unification with quarks}

\vspace{0.3cm}

 In this last Section, we try to apply to quarks the form of mass matrix  
which was worked out above for leptons. To this end, we conjecture for three 
generations of up quarks $ u\,,\,c\,,\,t $ and down quarks $ d\,,\,s\,,\,b $
the mass matrices $\left(M_{i\,j}^{(u)}\right)$ and $\left(M_{i\,j}^{(d)}
\right)$, respectively, essentially of the form (1), where the label $ f = u\,,
\,d $ denotes now up and down quarks. The only modification introduced is a 
new real constant $ C^{(f)}$ added to $\varepsilon^{(f)}$ in the element
$ M^{(f)}_{33}$ which now reads

\begin{equation}
M^{(f)}_{33} = \frac{24 \mu^{(f)}}{25\cdot 29}\left(624 + \varepsilon^{(f)} 
+ C^{(f)}\right)\;.
\end{equation}

 Since for quarks the mass scales $\mu^{(u)}$ and $\mu^{(d)}$ are expected to 
be even more important than the scale $\mu^{(e)}$ for charged leptons, we 
assume that the off--diagonal elements of mass matrices $\left(M_{i\,j}^{(u)}
\right)$ and $\left(M_{i\,j}^{(d)}\right)$ can be considered as a small 
perturbation of their diagonal terms. Then, in the lowest perturbative order, 
we obtain the following mass formulae


\begin{eqnarray}
m_{u,d} & = & \frac{\mu^{(u,d)} }{29} \varepsilon^{(u,d)} - A^{(u,d)}
\left(\frac{\alpha^{(u,d)}}{\mu^{(u,d)}}\right)^2 \; , \nonumber \\ 
m_{c,s} & = & \frac{\mu^{(u,d)}}{29} \frac{4}{9}\left(80 + \varepsilon^{(u,d)}
\right) + \left( A^{(u,d)} - B^{(u,d)} \right)\left(\frac{\alpha^{(u,d)}}{
\mu^{(u,d)} }\right)^2 \; , \nonumber \\ 
m_{t,b} & = & \frac{\mu^{(u,d)}}{29} \frac{24}{25} \left(624 + \varepsilon^{(
u,d)} + C^{(u,d)} \right) + B^{(u,d)}\left(\frac{\alpha^{(u,d)}}{\mu^{(u,d
)}}\right)^2 \;,
\end{eqnarray}


\ni where

\begin{equation}
A^{(u,d)} = \frac{\mu^{(u,d)}}{29}\,\frac{36}{320 - 5\varepsilon^{(u,d)} }\;
\;,\;\; B^{(u,d)} = \frac{\mu^{(u,d)}}{29}\,\frac{10800}{31696 + 54 C^{(u,d)}
+ 29\varepsilon^{(u,d)}}\;.
\end{equation}

\ni In Eqs. (35), the relative smallness of perturbating terms is more pronoun%
ced due to extra factors. In our discussion, we will take for experimental 
quark masses the arithmetic means of their lower and upper limits quoted in the
Review of Particle Physics [3] {\it i.e.},


\begin{equation}
m_u = 3.3 \,{\rm MeV}\;,\; m_c = 1.3 \,{\rm GeV}\;,\;m_t = 174 \,{\rm GeV}
\end{equation}

\ni and


\begin{equation}
m_d = 6 \,{\rm MeV}\;,\; m_s = 120 \,{\rm MeV}\;,\;m_b = 4.3 \,{\rm GeV}\;.
\end{equation}

 Eliminating from the unperturbed terms in Eqs. (35) the constants $\mu^{(u,d)}
$ and $\varepsilon^{(u,d)}$, we derive the correlating formulae being counter%
parts of Eqs. (2) for charged leptons:

\begin{eqnarray}
m_{t,b} & = & \frac{6}{125} \left( 351 m_{c,s} - 136 m_{u,d} \right) + 
\frac{\mu^{(u,d)}}{29}\frac{24}{25} C^{(u,d)} \nonumber \\ & & - \frac{1}{125} 
\left(2922 A^{(u,d)} - 2231 B^{(u,d)}\right) \left(\frac{\alpha^{(u,d)}}{\mu^{(
u,d)}}\right)^2\;, \nonumber \\
\mu^{(u,d)} & = & \frac{29}{320} \left(9 m_{c,s} - 4m_{u,d}\right) -
\frac{29}{320} \left(5 A^{(u,d)} - 9 B^{(u,d)}\right) \left(\frac{\alpha^{(u,d
)}}{\mu^{(u,d)}}\right)^2 \;, \nonumber \\  
\varepsilon^{(u,d)} & = & \frac{29 m_{u,d}}{\mu^{(u,d)}} + \frac{29}{\mu^{
(u,d)}} A^{(u,d)} \left(\frac{\alpha^{(u,d)}}{\mu^{(u,d)}}\right)^2 \;. 
\end{eqnarray}

\ni The unperturbed parts of these relations are:

\begin{eqnarray}
\stackrel{\circ}{m}_{t,b} & = & \frac{6}{125} \left( 351 m_{c,s} - 
136 m_{u,d} \right) + \frac{\stackrel{\circ}{\mu}^{(u,d)}}{29}\frac{24}{25} 
\stackrel{\circ}{C}^{(u,d)} \nonumber \\ & = & \left\{\begin{array}{c} 
21.9\\1.98 \end{array}\right\}\,{\rm GeV} + \frac{\stackrel{\circ}{\mu}^{(u,
d)}}{29}\frac{24}{25} \stackrel{\circ}{C}^{(u,d)} \;, \nonumber \\
\stackrel{\circ}{\mu}^{(u,d)} & = & \frac{29}{320} \left(9 m_{c,s} - 4m_{u,d}
\right) = \left\{\begin{array}{c} 1060 \\ 95.7 \end{array}\right\}\,{\rm MeV} 
\;, \nonumber \\  
\stackrel{\circ}{\varepsilon}^{(u,d)} & = & \frac{29 m_{u,d}}{\stackrel{
\circ}{\mu}^{(u,d)}} = \left\{\begin{array}{l} 0.0904 \\ 1.82 \end{array}
\right\} \;.
\end{eqnarray}

\ni In the spirit of our perturbative approach, the "coupling" constant $
\alpha^{(u,d)}$ can be put zero in all perturbing terms in Eqs. (35) and (39), 
except for $\alpha^{(u,d)\,2}$ in the numerator of the factor $(\alpha^{(u,d)}
/\mu^{(u,d)})^2$ that now becomes $(\alpha^{(u,d)}/\stackrel{\circ}{\mu}^{(u,d
)})^2$. Then, $ A^{(u,d)}$ and $ B^{(u,d)}$ are replaced by 

\begin{equation}
\stackrel{\circ}{A}^{(u,d)} = \frac{\stackrel{\circ}{\mu}^{(u,d)}}{29}
\frac{36}{320 - 5\stackrel{\circ}{\varepsilon}^{(u,d)} }\;
\;,\;\; \stackrel{\circ}{B}^{(u,d)} = \frac{\stackrel{\circ}{\mu}^{(u,d)}}{29} 
\frac{10800}{31696 + 54 \stackrel{\circ}{C}^{(u,d)} + 29\stackrel{\circ}{
\varepsilon}^{(u,d)}}\;.
\end{equation}

\ni Note that the first Eq. (35) can be rewritten identically as $m_{u,d} =
\;\stackrel{\circ}{\mu}^{(u,d)}\,\stackrel{\circ}{\varepsilon}^{(u,d)}\!\!\!/
29 $ according to the third Eq. (40).

 We shall be able to return to the discussion of quark masses after the estim%
ation of constants $\alpha^{(u)}$ and $\alpha^{(d)}$ is made. Then, we shall 
determine the parameters $ C^{(u)}$ and $C^{(d)}$ (as well as their unperturbed
parts $\stackrel{\circ}{C}^{(u)}$ and $\stackrel{\circ}{C}^{(d)}$) playing here
an essential role in providing large values for $ m_t $ and $ m_b $.

 At present, we find the unitary matrices $\left(U_{ij}^{(u,d)}\right)$ that 
diagonalize the mass matrices $\left(M_{ij}^{(u,d)}\right)$ according to the 
relations $ U^{(u,d)\,\dagger}M^{(u,d)}U^{(u,d)} = $ diag$(m_{u,d}\,,\,
m_{c,s}\,,\,m_{t,b})$. In the lowest perturbative order, the result has the 
form (4) with the necessary replacement of labels:


\begin{equation}
(e) \rightarrow (u)\;\;{\rm or}\;\;(d)\;,\;\mu \rightarrow c\;\;{\rm or}\;\;s
\;,\;\tau \rightarrow t\;\;{\rm or}\;\;b\;,
\end{equation}

\ni respectively.

 Then, the elements $ V_{ij}$ of the \CKM matrix $ V = U^{(u)\,\dagger}U^{(d)}$ 
can be calculated with the use of Eqs. (42) in the lowest perturbative order. 
Six resulting off--diagonal elements are:


\begin{eqnarray}
V_{us} & = & -V^*_{cd} = \frac{2}{29}\left(\frac{\alpha^{(d)}}{m_s} 
e^{i\varphi^{(d)}} - \frac{\alpha^{(u)}}{m_c} e^{i\varphi^{(u)}} \right) \;,
\nonumber \\ 
V_{cb} & = & -V^*_{ts} = \frac{8\sqrt{3}}{29}\left(\frac{\alpha^{(d)}}{m_b} 
e^{i\varphi^{(d)}} - \frac{\alpha^{(u)}}{m_t} e^{i\varphi^{(u)}} \right) \simeq
\frac{8\sqrt{3}}{29} \frac{\alpha^{(d)}}{m_b} e^{i\varphi^{(d)}} \;, 
\nonumber \\
V_{ub} & \simeq & -\frac{16\sqrt{3}}{841}\frac{\alpha^{(u)}\alpha^{(d)} }{
m_c m_b} e^{i(\varphi^{(u)}+\varphi^{(d)})} \;, \nonumber \\ 
V_{td} & \simeq & \frac{16\sqrt{3}}{841} 
\frac{\alpha^{(d)\,2}}{m_s m_b}\,e^{-2i\varphi^{(d)}} \;,
\end{eqnarray}

\ni where the indicated approximate steps were made due to the inequality $ m_t
\gg m_b $ and/or under the assumption that $\alpha^{(u)}/m_c \gg \alpha^{(d)}
/m_b $ [{\it cf.} the conjecture (46)]. All three diagonal elements are real 
and positive in a good approximation:

\begin{equation}
V_{ud} \simeq 1 - \frac{1}{2}|V_{us}|^2\;,\;V_{cs} \simeq 1 - \frac{1}{2}
|V_{us}|^2 - \frac{1}{2}|V_{cb}|^2\;,\;V_{tb} \simeq 1 - \frac{1}{2}|V_{cb}|^2
\;.
\end{equation}

\ni In fact, in the lowest perturbative order,


\begin{equation}
\arg V_{ud} \simeq \frac{4}{841} \frac{\alpha^{(u)}\alpha^{(d)}}{m_c m_s}
\sin \left(\varphi^{(u)} - \varphi^{(d)}\right)\frac{180^\circ}{\pi} \simeq 
-\arg V_{cs}\;,\;\arg V_{tb} \simeq 0 \;,
\end{equation}

\ni what gives $ \arg V_{ud} = 0.88^\circ = -\arg V_{cs}$, if the values (46), 
(49) and (52) are used.

 Taking as an input the experimental value $|V_{cb}| = 0.0395 \pm 0.0017 $ [3],
we estimate from the second Eq. (43) that 

\begin{equation}
\alpha^{(d)} \simeq \frac{29}{8\sqrt{3}}\, m_b\, |V_{cb}| = (355 \pm 15)\;{\rm 
MeV} \;,
\end{equation}

\ni where $ m_b = 4.3 $ GeV. In order to estimate also $\alpha^{(u)}$, we will
tentatively conjecture the approximate proportion

\begin{equation}
\alpha^{(u)} : \alpha^{(d)} \simeq Q^{(u)\,2} : Q^{(d)\,2} = 4
\end{equation}

\ni to hold, where $ Q^{(u)} = 2/3 $ and $ Q^{(d)} = -1/3 $ are quark electric 
charges. Note that in the case of leptons we had $\alpha^{(\nu)} : \alpha^{(e)}
= 0.37 : (\sqrt{180}\,\times 10^6) = 2.8\times 10^{-8}$ for the central value of
$\alpha^{(e)}$ [{\it cf.} Eqs. (3) and (30)], what is consistent with the 
analogical approximate proportion

\begin{equation}
\alpha^{(\nu)} : \alpha^{(e)} \simeq Q^{(\nu)\,2} : Q^{(e)\,2} = 0 \;,
\end{equation}

\ni where $Q^{(\nu)} = 0 $ and $ Q^{(e)} = -1 $ are lepton electric charges. 
Under the conjecture (47):

\begin{equation}
\alpha^{(u)} \simeq (1420 \pm 60)\, {\rm MeV} \;.
\end{equation}

\ni In this case, from the second and third Eq. (43) we obtain the prediction

\begin{equation}
|V_{ub}|/|V_{cb}| \simeq \frac{2}{29}\frac{\alpha^{(u)}}{m_c} \simeq 0.0753 
\pm 0.0032 \;,
\end{equation}

\ni where $ m_c = 1.3 $ GeV. This is consistent with the experimental 
figure $|V_{ub}|/|V_{cb}| = 0.08 \pm 0.02 $ [3].

 Now, with the experimental value $|V_{us}| = 0.2196 \pm 0.0023$ [3] as 
another input, we can calculate from the first Eq. (43) the phase difference 
$\varphi^{(u)} - \varphi^{(d)}$. In fact, taking the absolute value of this 
equation, we get

\begin{equation}
\cos\left(\varphi^{(u)} - \varphi^{(d)}\right) = \frac{1}{8}\frac{m_c}{m_s}
\left[1 + 16\left(\frac{m_s}{m_c}\right)^2 - \frac{841}{4}\left(\frac{m_c}{
\alpha^{(d)}}\right)^2 |V_{us}|^2 \right] = - 0.0301 
\end{equation}

\ni with $ m_c = 1.3 $ GeV  and $ m_s = 120 $ MeV, if the proportion (47) is 
taken into account. Here, the central values of $\alpha^{(d)}$ and $|V_{us}|$ 
were used. Hence,

\begin{equation}
\varphi^{(u)} - \varphi^{(d)} = 91.7^\circ = -88.3^\circ + 180^\circ
\end{equation}

\ni so, this phase difference turns out to be near $ 90^\circ $. Then, 
calculating the argument of the first Eq. (43), we infer that

\begin{equation}
\tan\left(\arg V_{us} - \varphi^{(d)}\right) = -4 \,\frac{m_s}{m_c}\,
\frac{\sin\left(\varphi^{(u)} - \varphi^{(d)}\right)}{1 - 4 ({m_s}/{m_c}) 
\cos\left(\varphi^{(u)} - \varphi^{(d)}\right)} = - 0.365 \;,
\end{equation}


\ni what gives


\begin{equation}
\arg V_{us} = -20.1^\circ + \varphi^{(d)} \;.
\end{equation}

 The results (52) and (54) together with the formula (43) enable us to evaluate
the rephasing--invariant CP--violating phases

\begin{equation}
\arg (V_{us}^*V_{cb}^*V_{ub}) = 20.1^\circ - 88.3^\circ = -68.2^\circ 
\end{equation}


\ni and 


\begin{equation}
\arg (V_{cd}^*V_{ts}^*V_{td}) = -20.1^\circ \;,
\end{equation}

\ni which turn out to be near to -70$^\circ$ and -20$^\circ$, respectively
(they are invariant under quark rephasing equal for up and down quarks of the 
same generation). Note that the sum of arguments (55) and (56) is always equal 
to $\varphi^{(u)} - \varphi^{(d)} - 180^\circ $. Carrying out quark rephasing 
(equal for up and down quarks of the same generation), where

\begin{equation}
\arg V_{us} \rightarrow 0 \;,\; \arg V_{cb} \rightarrow 0 \;,\; \arg V_{cd}
\rightarrow 180^\circ \;,\; \arg V_{ts} \rightarrow 180^\circ
\end{equation}

\ni and $\arg V_{ud}$, $\arg V_{cs}$, $\arg V_{tb}$ remain unchanged, we
conclude from Eqs. (55) and (56) that

\begin{equation}
\arg V_{ub} \rightarrow -68.2^\circ \;,\; \arg V_{td} \rightarrow -20.1^\circ
\;.
\end{equation}

\ni The sum of arguments (58) after rephasing (57) is always equal to $
\varphi^{(u)} - \varphi^{(d)} - 180^\circ $.

 Thus, in this quark phasing, we predict the following \CKM matrix:

\begin{equation}
\left( V_{ij} \right) = \left(\begin{array}{ccc}
0.976 & 0.220 & 0.00297\,e^{-i\,68.2^\circ}\\ -0.220 & 0.975 & 0.0395 \\  
0.00805\,e^{-i\,20.1^\circ} & -0.0395 & 0.999 \end{array}\right)\;.
\end{equation}

\ni Here, only $|V_{us}|$ and $|V_{cb}|$ [and quark masses $m_s\,,\;m_c\,,\;
m_b $ consistent with the mass matrices $\left( M^{(u)}_{ij}\right)$ and 
$\left( M^{(d)}_{ij}\right)$] are our inputs, while all other matrix elements $
V_{ij}$, partly induced by unitarity, are evaluated from the relations derived 
in this Section from the Hermitian mass matrices $\left( M^{(u)}_{ij}\right)$ 
and $\left( M^{(d)}_{ij}\right)$ [and the conjectured proportion (47)]. The 
independent predictions are $|V_{ub}|$ and arg$ V_{ub}$. In Eq. (59), the small
phases arising from Eqs. (45), $\arg V_{ud} = 0.9^\circ$ and $\arg V_{cs} = -
0.9^\circ$, are neglected (here, arg $(V_{ud}V_{cs}V_{tb}) = 0$).

 The above prediction of $ V_{ij}$ implies the following values of Wolfenstein 
parameters [3]:

\begin{equation}
\lambda = 0.2196\;\;,\;\;A = 0.819 \;\;,\;\;\rho = 0.127\;\;,\;\;\eta = 0.319 
\end{equation}

\ni and of unitary--triangle angles:

\begin{equation}
\gamma = \arctan \frac{\eta}{\rho} = - \arg V_{ub} = 68.2^\circ\;\;,\;\;\beta
= \arctan \frac{\eta}{1-\rho} = - \arg V_{td} = 20.1^\circ \;.
\end{equation}

\ni The predicted large value of $\gamma $ follows the present experimental 
tendency.

 If instead of the central value $|V_{us}| = 0.2196$ we take as the input 
the range $|V_{us}| = 0.2173$ to 0.2219, we obtain from Eq. (51) $\varphi^{(u)}
- \varphi^{(d)} = 89.8^\circ\;\;{\rm to}\;\;93.6^\circ$ (with $|V_{cb}| = 
0.0395$ giving $\alpha^{(d)} = 355$ MeV), what implies through Eq. (53) that 
arg $V_{us}- \varphi^{(d)} = {-20.3}^\circ\;\;{\rm to}\;\; {-19.8}^\circ $. 
Then, after rephasing (57), arg$V_{ub} = -69.9^\circ\;\;{\rm to}\;\;-66.6^\circ
$ and ${\arg V_{td}} = -20.3^\circ\;\;{\rm to}\;\;{-19.8}^\circ$. In this case,
the Wolfenstein parameters are $\lambda = 0.2173 $ to 0.2219, $ A = 0.837 $ to 
0.802, $\rho = 0.119 $ to 0.135 and $\eta = 0.325 $ to 0.312 (here, $\lambda
\sqrt{\rho^2 + \eta^2} = |V_{ub}|/|V_{cb}| = 0.0753 $ is fixed). Thus, $\gamma 
= - \arg V_{ub} = 69.9^\circ\;\;{\rm to}\;\;66.6^\circ$ and $\beta = - \arg 
V_{td} = 20.3^\circ \;\;{\rm to}\;\;19.8^\circ$.

 In contrast, if the central value $|V_{cd}| = 0.0395$ (giving $\alpha^{(d)} = 
355$ MeV) is replaced by the input of the range $V_{cd} = 0.0378$ to 0.0412 
(corresponding to $\alpha^{(d)} = 340$ to 370 MeV), we calculate from Eq. (51) 
that $\varphi^{(u)} - \varphi^{(d)} = 97.3^\circ\;\;{\rm to}\;\;84.9^\circ$ 
(with $|V_{us}| = 0.2196 $), what leads to arg$V_{us} - \varphi^{(d)} = -
19.3^\circ\;\;{\rm to}\;\;-20.9^\circ$. Hence, after rephasing (57), arg$V_{ub}
= -63.4^\circ\;\;{\rm to}\;\;{-74.6^\circ}$ and ${\arg V_{td}} = -19.3^\circ\;\;
{\rm to}\;\;-20.9^\circ$. In this case, the Wolfenstein parameters take the 
values $\lambda = 0.2196 $, $A = 0.784 $ to 0.854, $\rho = 0.149$ to 0.0951 and
$\eta = 0.298 $ to 0.345. Thus, $\gamma = - \arg V_{ub} = 63.4^\circ\;\;{\rm 
to} \;\;74.6^\circ$ and $\beta = - \arg V_{td} = 19.3^\circ \;\;{\rm to}\;\;
20.9^\circ$. Here, $|V_{ub}| = 0.00273 $ to 0.00323 and $|V_{td}| = 0.00738 $ 
to 0.00874.

  Eventually, we may turn back to quark masses. From the third Eq. (35) we can 
evaluate

\begin{equation}
C^{(u,d)} = \frac{29}{\mu^{(u,d)}}\,\frac{25}{24}\,m_{t,b} - 624 - 
\varepsilon^{(u,d)} - \frac{29}{\mu^{(u,d)}}\,\frac{25}{24}\,B^{(u,d)}\left(\frac
{\alpha^{(u,d)}}{\mu^{(u,d)}}\right)^2\;,
\end{equation}

\ni what, in the framework of our perturbative approach, gives

\begin{eqnarray}
C^{(u,d)} & = & \stackrel{\circ}{C}^{(u,d)} + \frac{29}{\stackrel{\circ}{\mu}^{
(u,d)}}\, \frac{25}{24}\,m_{t,b} \,\frac{29}{320\stackrel{\circ}{\mu}^{(u,d)}}
\,\left(5\stackrel{\circ}{A}^{(u,d)} - 9\stackrel{\circ}{B}^{(u,d)}\right)\,
\left(\frac{\alpha^{(u,d)}}{\stackrel{\circ}{\mu}^{(u,d)}}\right)^2 \nonumber
\\ & & - \frac{29}{\stackrel{\circ}{\mu}^{(u,d)}}\,\left(
\stackrel{\circ}{A}^{(u,d)} + \stackrel{\circ}{B}^{(u,d)}\right)\,\left(
\frac{\alpha^{(u,d)}}{\stackrel{\circ}{\mu}^{(u,d)}}\right)^2\;,
\end{eqnarray}

\ni where

\begin{equation}
\stackrel{\circ}{C}^{(u,d)} = \frac{29}{\stackrel{\circ}{\mu}^{(u,d)}}\, 
\frac{25}{24}\,m_{t,b} - 624 - \stackrel{\circ}{\varepsilon}^{(u,d)} =
\left\{\begin{array}{c} 4339 \\ 733.2 \end{array}\right\} =
\left\{\begin{array}{r} 4340 \\ 733 \end{array}\right\} \;.
\end{equation}

\ni With the central values of $\alpha^{(u)}$ and $\alpha^{(d)}$ as estimated 
in Eqs. (46) and (49) we find from Eqs. (41)

\begin{equation}
\stackrel{\circ}{A}^{(u,d)}\left(\frac{\alpha^{(u,d)}}{\stackrel{\circ}{\mu
}^{(u,d)}}\right)^2 = \left\{\begin{array}{r} 7.39 \\ 5.26 \end{array}
\right\}\,{\rm MeV}\;,\;
\stackrel{\circ}{B}^{(u,d)}\left(\frac{\alpha^{(u,d)}}{\stackrel{\circ}{\mu
}^{(u,d)}}\right)^2 = \left\{\begin{array}{r} 2.66 \\ 6.88 \end{array}
\right\}\,{\rm MeV} \;,
\end{equation}

\ni where

\begin{equation}
\frac{\stackrel{\circ}{\mu}^{(u,d)}}{29}\left(\frac{\alpha^{(u,d)}}{\stackrel{
\circ}{\mu}^{(u,d)}}\right)^2 = \left\{\begin{array}{r} 65.6 \\ 45.4 
\end{array}\right\}\,{\rm MeV}\;.
\end{equation}

\ni We calculate from Eqs. (63) with the use of values (65) that

\begin{equation}
C^{(u,d)} = \left\{\begin{array}{c} 4339 + 5.25 \\ 733.2 - 49.5 \end{array}
\right\} = \left\{\begin{array}{c} 4344 \\ 683.7 \end{array}
\right\}  = \left\{\begin{array}{r} 4340 \\ 684 \end{array}
\right\} \;.
\end{equation}

 Similarly, from the second Eq. (39), making use of the values (65), we obtain

\begin{equation}
{\mu}^{(u,d)} = \left\{\begin{array}{c} 1060 - 1.18 \\ 95.7 + 3.23 \end{array}
\right\}\,{\rm MeV} = \left\{\begin{array}{c} 1059 \\ 98.9 \end{array}\right\}
\,{\rm MeV} = \left\{\begin{array}{c} 1060 \\ 98.9 \end{array}\right\}
\,{\rm MeV}\;.
\end{equation}

 We can easily check that, with the values (40) for $\stackrel{\circ}{\mu}^{(u,
d)}$ and $\stackrel{\circ}{\varepsilon}^{(u,d)}$ and the value (64) for $
\stackrel{\circ}{C}^{(u,d)}$ determined as above from quark masses, the 
unperturbed parts of mass formulae (35) reproduce correctly these masses. In 
fact,

\begin{eqnarray}
\stackrel{\circ}{m}_{u,d} & = &  \frac{\stackrel{\circ}{\mu}^{(u,d)}}{29}\,
\stackrel{\circ}{\varepsilon}^{(u,d)} = \left\{\begin{array}{c} 3.3 \\ 6 
\end{array}\right\}\,{\rm MeV}\;, \nonumber \\
\stackrel{\circ}{m}_{c,s} & = &  \frac{\stackrel{\circ}{\mu}^{(u,d)}}{29}
\,\frac{4}{9}\left(80 + \stackrel{\circ}{\varepsilon}^{(u,d)}\right) =
\left\{\begin{array}{r} 1300 \\ 120 \end{array}\right\}\,{\rm MeV}\;, 
\nonumber \\
\stackrel{\circ}{m}_{t,b} & = &  \frac{\stackrel{\circ}{\mu}^{(u,d)}}{29}
\,\frac{24}{25}\left(624 + \stackrel{\circ}{\varepsilon}^{(u,d)} + 
\stackrel{\circ}{C}^{(u,d)}\right) = \left\{\begin{array}{c} 174 \\ 4.3 
\end{array}\right\}\,{\rm GeV}\;.
\end{eqnarray}

\ni The same is true for the unperturbed part of the first correlating formula
(39). The --- here omitted --- corrections to Eqs. (69), arising from all pert%
urbing terms in the mass formulae (35) (including the corrections from ${\delta
\mu^{(u,d)}} $, $\delta \varepsilon^{(u,d)}$ and $\delta C^{(u,d)}$), are 
relatively small, {\it viz.}

\begin{equation}
\delta m_{u,d}\! = \!\left\{\begin{array}{r}3.7\times 10^{-3} \\ -2.0\times 
10^{-1}\end{array}\right\}\,{\rm MeV}\,,\,\delta m_{c,s}\! = \!\left\{
\begin{array}{r}9.5 \\ -3.8 \end{array}\right\}\,{\rm MeV}\,,\,\delta m_{t,b}\!
= \!\left\{\begin{array}{r} 170 \\ -74 \end{array}\right\}\,{\rm MeV}\,, 
\end{equation}

\ni respectively.

We would like to stress that, in contrast to the case of charged leptons, where
$m_\tau $ has been predicted from $m_e $ and $m_\mu $, in the case of up and 
down quarks two extra parameters $ C^{(u)}$ and $ C^{(d)}$ appear necessarily 
to provide large masses $m_t $ and $m_b $ (much larger than $m_\tau $). They 
cause that $m_t $ ($m_b $) cannot be predicted from $m_u $ and $m_c $ ($m_d $ 
and $m_s $), till the new parameters are quantitatively understood.

 Note that a conjecture about $ C^{(u)}$ and $ C^{(d)}$ might lead to a predic%
tion for quark masses and so, introduce changes in the "experimental" quark 
masses (37) and (38) accepted here. The same is true for a conjecture about $ 
\varphi^{(u)}$ and $\varphi^{(d)}$.

 For instance, the conjecture that the phase difference $\varphi^{(u)} - 
\varphi^{(d)}$ is maximal,

\begin{equation}
\varphi^{(u)} - \varphi^{(d)} = 90^\circ \;,
\end{equation}

\ni leads through the first equality in Eq. (51) to the condition

\begin{equation}
1 + 16\left(\frac{m_s}{m_c}\right)^2 - \frac{841}{4}\left(\frac{m_s}{
\alpha^{(d)}}\right)^2 |V_{us}|^2 = 0
\end{equation}

\ni predicting for $ s $ quark the mass

\begin{equation}
m_s = 118.7\,{\rm MeV} = 119\,{\rm MeV} 
\end{equation}

\ni (with $\alpha^{(d)} = 355 $ MeV), being only slightly lower than the value 
120 MeV used previously. Here, $ m_c $ and $ m_b $ are kept equal to 1.3 and 4.3
GeV, respectively (also masses of $ u\,,\,d $ and $t $ quarks are not changed, 
while $\stackrel{\circ}{\mu}^{(d)}$, $\stackrel{\circ}{\varepsilon}^{(d)}$ and 
$\stackrel{\circ}{C}^{(d)}$ change slightly). Then, from the first equality in 
Eq. (53)

\begin{equation}
\tan\left(\arg V_{us} - \varphi^{(d)}\right) = -4 \,\frac{m_s}{m_c}
= - 0.365 \;\;,\;\;\arg V_{us} = -20.1^\circ + \varphi^{(d)} \;.
\end{equation}

\ni After rephasing (57), this gives $\arg V_{ub} + \arg V_{td} = \varphi^{(u)}
- \varphi^{(d)} - 180^\circ = -90^\circ $, where

\begin{equation}
\arg V_{ub} = -69.9^\circ\;\;,\;\;\arg V_{td} = -20.1^\circ
\end{equation}

\ni {\it i.e.}, practically $-70^\circ$ and $-20^\circ$. All $ |V_{ij}|$ remain
unchanged (with our inputs of $ |V_{us}| = 0.2196 $ and $ |V_{cb}| = 0.0395 $),
except for $ |V_{td}| $ which changes slightly, becoming

\begin{equation}
|V_{td}|  = 0.00814\;.
\end{equation}

\ni Thus, in the \CKM matrix predicted in Eq. (59), only $ |V_{td}| $ and the
phases (75) show some changes. The Wolfenstein parameters are

\begin{equation}
\rho = 0.118 \;\;,\;\; \eta = 0.322
\end{equation}

\ni and $\lambda $ and $ A $ unchanged (here, the sum $\rho^2 + \eta^2 = 0.118$
is also unchanged). Hence, $\gamma + \beta = 90^\circ $ and $\alpha = 180^\circ
- \gamma - \beta = 90^\circ $, where

\begin{equation}
\gamma = \arctan \frac{\eta}{\rho} = - \arg V_{ub} = 69.9^\circ \;\;,\;\;
\beta = \arctan \frac{\eta}{1 - \rho} = - \arg V_{td} = 20.1^\circ\;.
\end{equation}

\ni So, in the case of conjecture (71), the new restrictive relation

\begin{equation}
\frac{\eta}{\rho} = \frac{1 - \rho}{\eta}\;\;,\;\; \rho^2 + \eta^2 = \rho
\end{equation}

\ni holds, implying the prediction

\begin{equation}
|V_{td}| /|V_{ub}| = \sqrt{\frac{(1-\rho)^2 + \eta^2}{\rho^2 + \eta^2}} =
\frac{\eta}{\rho} = 2.74 \;,
\end{equation}

\ni due to the definition of $\rho $ and $\eta $ from $ V_{ub}$ and $ V_{td}$.
It is in agreement with our figures for $ |V_{td}|$ and $ |V_{ub}|$. Then, the 
new relationship

\begin{equation}
\frac{1}{4}\frac{m_c}{m_s}= \frac{\alpha^{(d)} m_c}{\alpha^{(u)} m_s} = 
\frac{\eta}{\rho} 
\end{equation}

\ni follows for quark masses $m_c $, $m_s $ and Wolfenstein parameters $\rho $,
$\eta $, in consequence of Eqs. (43) and the conjectured proportion (47). 
Both its sides are really equal for our values of $m_c $, $m_s$ and $\rho $,
$\eta $.

 Thus, summarizing, we cannot predict quark masses without an {\it additional}
knowledge or conjecture about the constants $\mu^{(u,d)}$, $\varepsilon^{(u,d)
}$, $ C^{(u,d)}$, $\alpha^{(u,d)}$ and $\varphi^{(u,d)}$ (in particular, the
conjecture (71) predicting $ m_s $ may be natural). However, we always describe
them correctly. If we describe them {\it jointly} with quark mixing parameters,
we obtain two independent predictions of $|V_{ub}|$ and $\gamma - \arg V_{ub}$:
~the whole \CKM matrix is calculated from the inputs of $|V_{us}|$ and of $|
V_{ub}| $ [and of quark masses $ m_s $, $ m_c $ and $ m_b $ consistent with 
the mass matrices $\left( M^{(u)}_{ij}\right) $ and $\left( M^{(d)}_{ij}\right)
 $].

 Concluding this Section, we can claim that our leptonic form of mass matrix 
works also in a promising way for up and down quarks. But, it turns out that, 
in the framework of the leptonic form of mass matrix, the heaviest quarks, $ t 
$ and $ b $, require an additional mechanism in order to produce the bulk of 
their masses (here, it is represented by the large constants $ C^{(u)}$ and 
$ C^{(d)}$). Such a mechanism, however, intervenes into the process of quark 
mixing only through quark masses, practically $m_t$ and $m_b$, and so, it does 
not modify for quarks the leptonic form of mixing mechanism.

\vfill\eject

~~~~
\vspace{0.6cm}

{\bf References}

\vspace{1.0cm}

{\everypar={\hangindent=0.5truecm}
\parindent=0pt\frenchspacing

{\everypar={\hangindent=0.5truecm}
\parindent=0pt\frenchspacing

1.~W.~Kr\'{o}likowski, in {\it Spinors, Twistors, Clifford Algebras and 
Quantum Deformations (Proc. 2nd Max Born Symposium 1992)}, eds. Z.~Oziewicz 
{\it et al.}, Kluwer Acad. Press, 1993; {\it Acta Phys. Pol.} {\bf B 27}, 
2121 (1996).

\vspace{0.15cm}

2.~W. Kr\'{o}likowski,{\it Acta Phys. Pol.} {\bf B 21}, 871 (1990); {\it 
Phys. Rev.} {\bf D 45}, 3222 (1992). 

\vspace{0.15cm}

3.{\it ~Review of Particle Physics}, {\it Eur. Phys. J.} {\bf C 3}, 1 (1998). 

\vspace{0.15cm}

4.~Y. Fukuda {\it et al.} (\UK Collaboration), {\it Phys. Rev. Lett.} {\bf 81},
1562 (1998); and references therein. 

\vspace{0.15cm}

5.~{\it Cf. e.g.}, J.N. Bahcall, P.I. Krastov and A.Y. Smirnow, hep--ph/%
9807216v2.

\vspace{0.15cm}

6.~C.~Athanassopoulos {\it et al.} (LSND Collaboration), {\it Phys. Rev.}
{\bf C 54} 2685 (1996); {\it Phys. Rev. Lett.} {\bf 77}, 3082 (1996); nucl--%
ex/9709006.

\vspace{0.15cm}

7.~M. Appolonio {\it et al.} (CHOOZ Collaboration), {\it Phys. Lett.} {\bf B 
420}, 397 (1998).

\vspace{0.15cm}

8.~W. Kr\'{o}likowski, hep--ph/9808207, to appear in {\it Acta Phys. Pol.} 
{\it B}.

\vfill\eject

\end{document}